\documentclass[aps,prl,preprint,superscriptaddress]{revtex4}

\usepackage{epsfig} 

\begin{document}


\title{Low ordered magnetic moment by off-diagonal frustration in undoped parent compounds to iron-based high-$T_c$ superconductors}

\author{J.P. Rodriguez}

\author{E.H. Rezayi}
\affiliation{Department of Physics and Astronomy, 
California State University, Los Angeles, California 90032}

\date{\today}

\begin{abstract}
A Heisenberg  model  over the square lattice
recently introduced by Si and Abrahams
to  describe local-moment magnetism in the new class of Fe-As
high-$T_c$ superconductors is analyzed
in the classical limit
and on a small cluster by exact diagonalization.
In the case of spin-1 iron atoms,
large enough Heisenberg exchange interactions between neighboring spin-1/2 moments 
on different iron $3d$ orbitals
that frustrate true magnetic order lead to
hidden magnetic order that violates Hund's rule.
It accounts for the low ordered magnetic moment observed by elastic neutron diffraction
in an undoped parent compound to Fe-As superconductors.
We predict that low-energy spin-wave excitations exist
at wavenumbers corresponding
to  either hidden N\' eel or  hidden ferromagnetic  order.
\end{abstract}

\maketitle

The recent discovery of a new class of high-$T_c$ superconductors that are
notably composed of iron-arsenic layers has reinvigorated the 
search for new superconductors\cite{new_sc}.
Iron  is usually detrimental to conventional superconductivity because
its magnetic moment breaks up Cooper pairs\cite{Anderson_theorem}.
Electronic conduction is confined primarily to the Fe-As layers in the new class
of high-$T_c$ superconductors, on the other hand\cite{wang}.
The nature of the magnetic moments in the iron atoms that make up the
new class of materials  may then be 
critical to the superconductivity that these systems display.

As in the copper-oxide high-$T_c$ superconductors,
the new Fe-As  superconductors are obtained by doping stoichiometric parent compounds.
Elastic neutron diffraction measurements on the  parent compound LaOFeAs reveal the
presence of  long-range spin-density wave (SDW) order at low temperature
that is commensurate with the square lattice of Fe atoms that make  up each layer\cite{delacruz}.
The magnetic moment associated with this collinear type of magnetic order
is only a fraction of the Bohr magneton, however.
Hund's rule is therefore  violated in the iron $3d$ orbitals of this new parent compound
for high-$T_c$ superconductivity.

In this Letter, we identify a route to low ordered magnetic moments in frustrated two-dimensional
magnets composed of local moments of spin one or higher.  
It is based on linear spin-wave analysis and exact diagonalization
of a  Heisenberg model over a square lattice of iron atoms 
that includes local Hund's rule coupling\cite{Si&A}.
We find that Heisenberg spin exchange between {\it different} $3d$ orbitals 
on neighboring iron atoms
leads to either  hidden N\' eel or hidden ferromagnetic order
if the exchange interaction is sufficiently frustrating.  
This may account for the low moment 
associated with collinear/SDW order
that is observed in LaOFeAs\cite{delacruz}.  
Low-energy spin-wave excitations are a natural consequence
of the hidden magnetic order, however.  
We predict that they
collapse to  the ground-state energy at the
respective N\' eel and ferromagnetic wave numbers.
Last, we identify a quantum phase transition into  hidden order
from a more familiar frustrated magnetic groundstate that obeys Hund's rule.

Recent transport measurements indicate that parent compounds to iron-based high-$T_c$ superconductors 
are bad insulators (metals)
close to a transition into a metallic (insulating) state\cite{wang}.
Further, classical spin-wave frequencies obtained from near-neighbor Heisenberg models 
can be used to fit the
measured spin-wave spectra in such parent compounds\cite{sw_exp}.
We believe, therefore, that a local-moment description of magentism in parent compounds to
iron-based high-$T_c$ superconductors 
is valid at low temperature.
Following Si and Abrahams, we then consider
a spin-1/2 Hamiltonian that 
contains near-neighbor Heisenberg exchange among local iron moments
within isolated layers plus Hund's-rule coupling\cite{Si&A}:
%
\begin{equation}
H = {1\over 2} J_0 \sum_i \biggl[\sum_{\alpha} {\bf S}_i (\alpha)\biggr]^2 +
    \sum_{\langle i,j \rangle} \sum_{\alpha , \beta} J_1^{\alpha,\beta}
                                      {\bf S}_i (\alpha) \cdot {\bf S}_j (\beta) +
    \sum_{\langle\langle i,j \rangle\rangle} \sum_{\alpha , \beta} J_2^{\alpha,\beta}
                                      {\bf S}_i (\alpha) \cdot {\bf S}_j (\beta).
\label{j0j1j2}
\end{equation}
Above, ${\bf S}_i(\alpha)$ is the spin operator that acts on the spin-1/2 state of orbital
$\alpha$ in the iron atom at site $i$.  The latter runs over the square lattice
of iron atoms that make up an isolated layer.  
The application of Hund's rule is controlled by a negative local
Heisenberg exchange constant $J_0 < 0$, 
while nearest neighbor and next-nearest neighbor Heisenberg exchange
across the links $\langle i,j\rangle$ and $\langle\langle i,j\rangle\rangle$
is controlled by the tensor exchange constants $J_1^{\alpha,\beta}$ and $J_2^{\alpha,\beta}$, respectively.
The strength of the crystal field at each iron atom compared to Hund's rule 
determines the number of orbitals per iron atom above. 
It can be as low as two for strong crystal fields,
and as high as four for weak crystal fields\cite{Si&A}.
We shall now search for groundstates of the $J_0$-$J_1$-$J_2$ model above (\ref{j0j1j2})
that exhibit low ordered magnetic moments that violate Hund's rule.  It is
useful to first consider the special case where all nearest-neighbor and next-nearest-neighbor
exchange coupling constants are equal, respectively: 
$J_1^{\alpha,\beta} = J_1$ and $J_2^{\alpha,\beta} = J_2$.
The Hamiltonian then reduces to 
$H = {1\over 2} J_0 \sum_i {\bf S}_i \cdot {\bf S}_i +
    J_1\sum_{\langle i,j \rangle} {\bf S}_i \cdot {\bf S}_j +
    J_2\sum_{\langle\langle i,j \rangle\rangle} {\bf S}_i \cdot {\bf S}_j$,
where ${\bf S}_i = \sum_{\alpha} {\bf S}_i(\alpha)$.
Observe now that ${\bf S}_i + {\bf S}_j$ commutes with ${\bf S}_i \cdot {\bf S}_i$,
and hence that the latter commutes with the Hamiltonian.  
This means that the total spin at a given
site $i$ is a good quantum number.  
The groundstate then obeys Hund's rule
in the classical limit
because states with
maximum total spin at a given site minimize both the Hund's-rule energy ($J_0 < 0$) and the
Heisenberg exchange energies in such a case.

\begin{figure}
\includegraphics[scale=0.70, angle=-90]{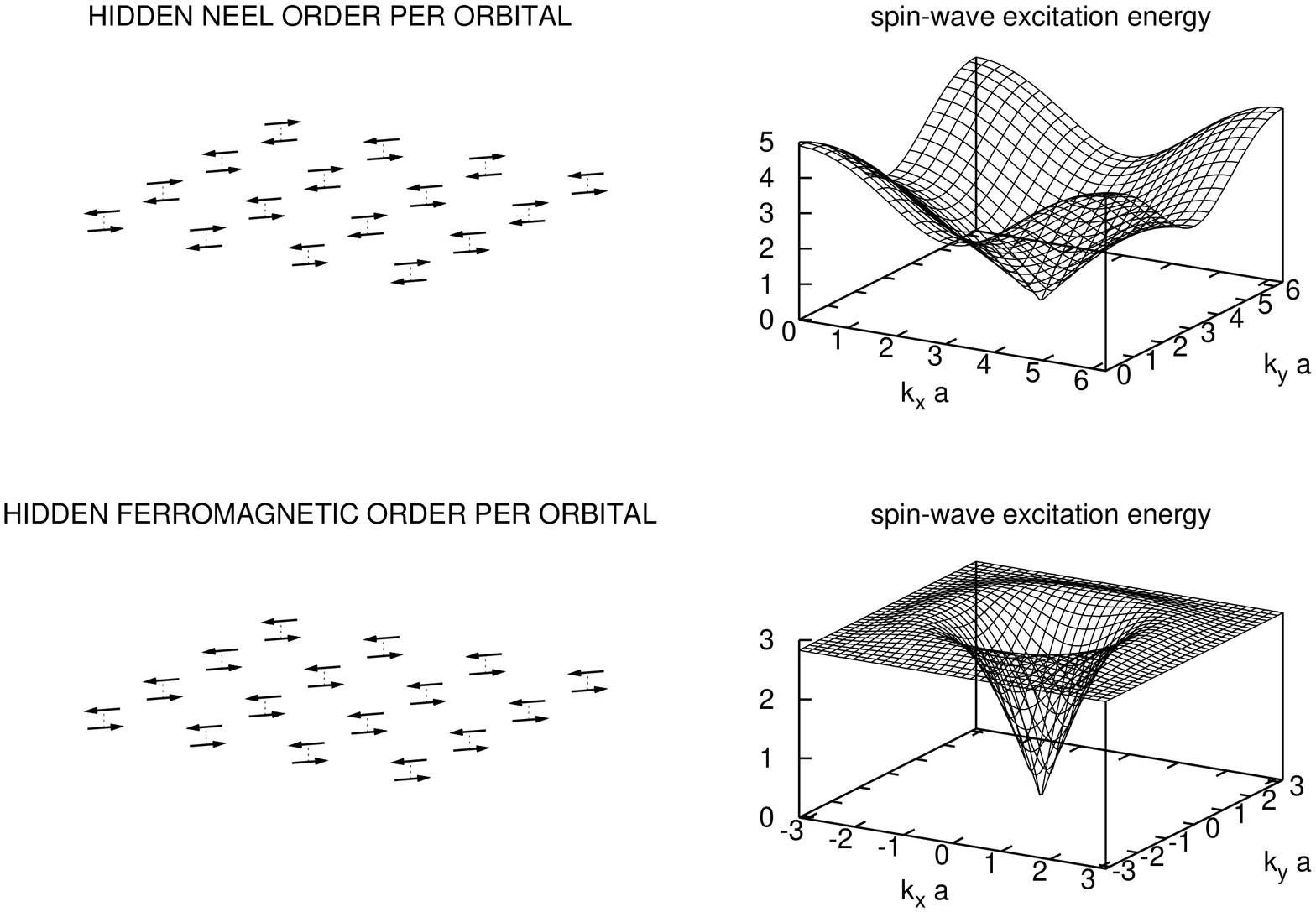}
\caption{The linear spin-wave spectrum for the Hamiltonian (\ref{j0j1j2})
is displayed in units of $|J_1|$
at off-diagonal $J_1 < 0$ and  $J_1 > 0$ respectively, 
at off-diagonal $J_2 = |J_1|/2$,
and with no Hund's rule coupling acting on two orbitals per site.
Hereafter, we set $\hbar\rightarrow 1$.}
\label{sw} 
\end{figure}

A violation of Hund's rule will therefore
require a  strong variation in the  Heisenberg exchange coupling
constants among the different iron orbitals.
This can be easily seen if we confine  ourselves
to the case of two $3d$-wave orbitals per site and choose
off-diagonal exchange coupling constants that lead to frustration
when Hund's rule is obeyed:
$J_1^{\alpha,\alpha} = 0 = J_2^{\alpha,\alpha}$,
while $J_1^{\alpha,\beta} = J_1$ and $J_2^{\alpha,\beta} = J_2$ if $\alpha \neq \beta$,
with  $J_2 > 0$.
In the limit of weak Hund's-rule coupling, $J_0\rightarrow 0$, 
the classical ground state {\it per orbital} is a
N\' eel state for $J_1 < 0$ and a ferromagnet for $J_1 > 0$. 
The spins at a given iron atom are equal and opposite across the two orbitals, however.
(See fig. \ref{sw}.) 
The moment associated with any type of 
true magnetic order must therefore vanish!
It is important to observe that the {\it hidden} magnetic order 
shown in fig. \ref{sw}
is stabilized by the addition of diagonal Heisenberg exchange coupling constants  that
are opposite in sign to the corresponding  off-diagonal ones.

Extremely low ordered moments are therefore possible at weak enough 
Hund's rule coupling, $J_0 < 0$,
when off-diagonal frustration exists:
$J_2^{\alpha, \beta} > 0$ at  $\alpha\neq\beta$.
The hidden order that is responsible for it is antiferromagnetic,
showing two sublattices (see fig. \ref{sw}).
Two spin-wave quanta per momentum $\hbar {\bf k}$ are then expected 
at $\hbar \omega_{\rm sw}$  above the groundstate energy\cite{af_sw_theory}.
Here, $\omega_{\rm sw}$ is the natural frequency, which is obtained by
linearizing the
dynamical equation
for classical precession by each  spin-1/2 moment,
$\dot {\bf S}_i (\alpha)  = {\bf S}_i (\alpha) \times \partial H / \partial {\bf S}_i (\alpha)$.
In the simple case where all diagonal Heisenberg exchange coupling constants are null,
it has the form
$\omega_{\rm sw} ({\bf k}) = (\Omega_+ \Omega_-)^{1/2}$,
with
\begin{eqnarray*}
\Omega_- &=& s|J_1| \sum_{n=x,y} (2\, {\rm sin}\, {1\over 2} k_n^{\prime} a )^2 +
     sJ_2 \sum_{n=+,-} (2\, {\rm sin}\, {1\over 2} k_n^{\prime} a)^2\\
\Omega_+ &=& 2sJ_0 + s|J_1| \sum_{n=x,y} (2\, {\rm cfn}\, {1\over 2} k_n^{\prime} a )^2 +
     sJ_2 \sum_{n=+,-} (2\, {\rm cos}\, {1\over 2} k_n^{\prime} a)^2 ,
\end{eqnarray*}
where ${\bf k}^\prime = {\bf k}$ or ${\bf k} - (\pi / a, \pi / a)$
and where ${\rm cfn} = {\rm cos}$ or ${\rm sin}$, respectively, 
in the case of hidden ferromagnetic order 
or  hidden N\' eel order per orbital,
at off-diagonal $J_1 > 0$ or  $J_1 < 0$. 
Above, $k_\pm^{\prime} = k_x^{\prime} \pm k_y^{\prime}$,
$a$ denotes the square lattice constant,
and $s$ is the electron spin.
Figure \ref{sw} depicts these spectra
at maximum off-diagonal frustration $J_2 = |J_1| / 2$.
The spin-wave velocity is then equal to
$c_{\rm sw} = s a [2 (|J_1| + 2 J_2) (J_0 +  4\, \theta (J_1) J_1 + 4 J_2)]^{1/2}$.
It collapses to zero at $J_0 = -4 (J_1+J_2)$ for off-diagonal $J_1 > 0$ and
 at $J_0 = -4 J_2$ for off-diagonal $J_1 < 0$, 
which serve as stability bounds for hidden ferromagnetic and N\' eel order, respectively.

The  above results indicate that large enough  
off-diagonal frustration in the $J_0$-$J_1$-$J_2$ model (\ref{j0j1j2})
induces a  quantum phase transition into hidden magnetic order that is unfrustrated, 
but that violates Hund's rule.  
We shall now study how the low-energy spectrum of states for the 
$J_0$-$J_1$-$J_2$ model  (\ref{j0j1j2}),
with two spin-1/2 moments per site,
evolves with the strength of the Hund's rule coupling
by applying the Lanczos technique numerically
on a 4 by 4 square lattice with periodic boundary conditions\cite{lanczos}.
As usual, we restrict the Hilbert space to states with equal numbers of up and down spins.
Next, translational invariance is exploited to reduce the Hamiltonian to block diagonal form, 
with each block labeled by a momentum quantum number.  
The allowed wave numbers, $(k_x a, k_y a)$ are then
$(0,0)$, $(\pi,0)$, ($\pi,\pi)$, $(\pi / 2, 0)$, $(\pi / 2, \pi / 2)$ and $(\pi, \pi / 2)$, 
plus their symmetric counterparts.  
The associated translational  symmetry reduces the dimension of each block to a little under
38,000,000 states.  
Spin-flip symmetry 
can then be exploited to further block-diagonalize the Hamiltonian at 
such momenta into two blocks
that are respectively even and odd under it.
The dimension of each of these subspaces is then a little under 19,000,000 states.
Each term in the Hamiltonian (\ref{j0j1j2}) permutes these Bloch-wave type states, 
and the permutations are stored in memory. 
Also, the value of the matrix element for Bloch waves 
that are composed of configurations of spin up and spin down
that display absolutely no non-trivial translation invariance is stored in
memory, while it is calculated otherwise.  This speeds up the application of the
Hamiltonian operator tremendously because
the vast majority of Bloch waves lie in the first category.  
The application of the Hamiltonian $H$ on a given state is
accelerated further by
enabling  shared-memory parallel computation through 
OpenMP directives.  
Last, we use the ARPACK subroutine library to apply the Lanczos technique
on the block-diagonal Hamiltonian operator just described\cite{arpack}.

\begin{figure}
\includegraphics[scale=0.70, angle=-90]{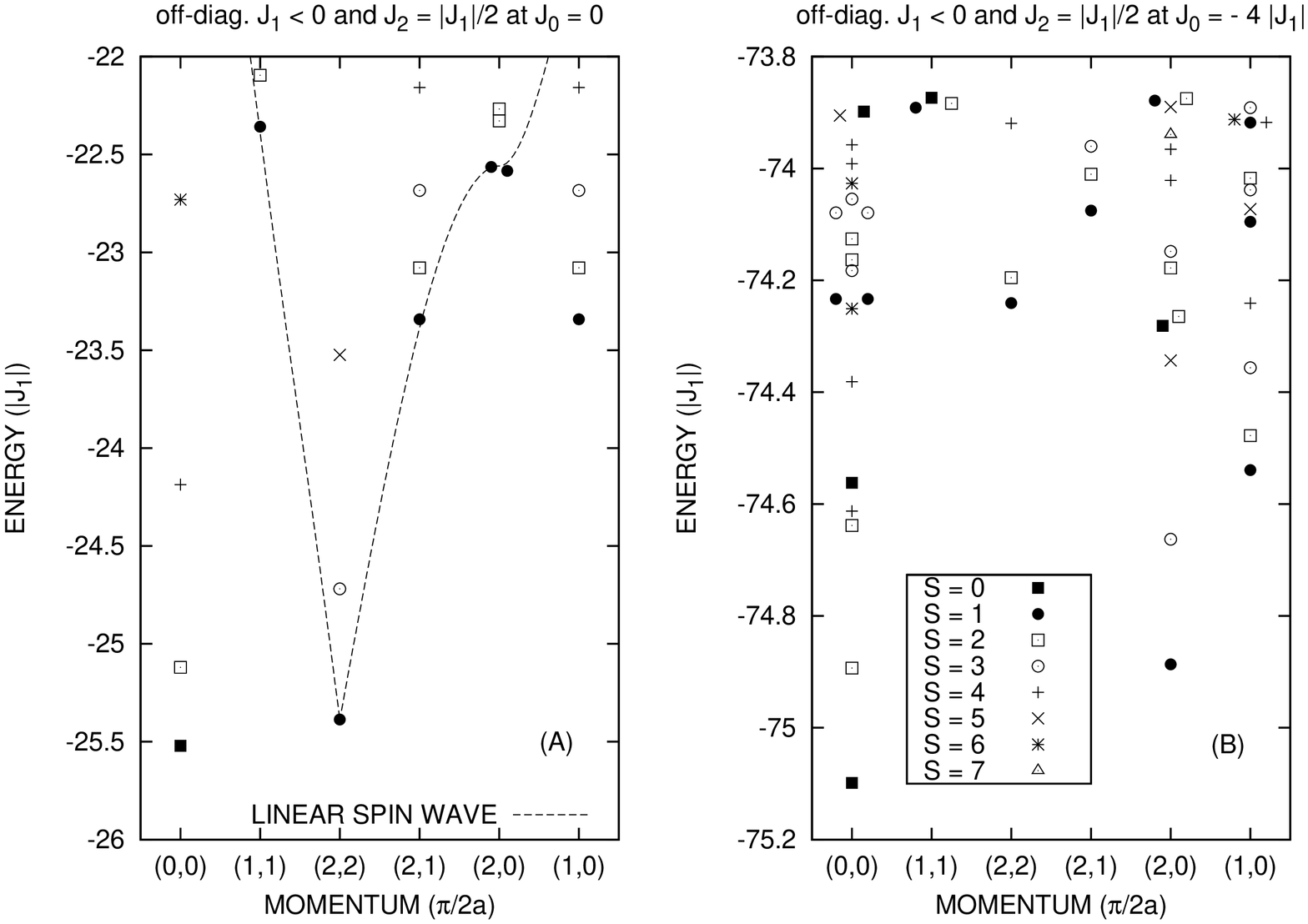}
\caption{Shown is the low-energy spectrum for $4\times 4\times 2$ spin-1/2 moments
that experience off-diagonal ferromagnetic and frustrating
Heisenberg exchange at weak and at moderately strong Hund's rule coupling.
The lowest-energy spin-1 state at momentum $(\pi,\pi)$ is used as the reference
for the linear spin-wave approximation.}
\label{fe+_spctra}
\end{figure}

\begin{figure}
\includegraphics[scale=0.70, angle=-90]{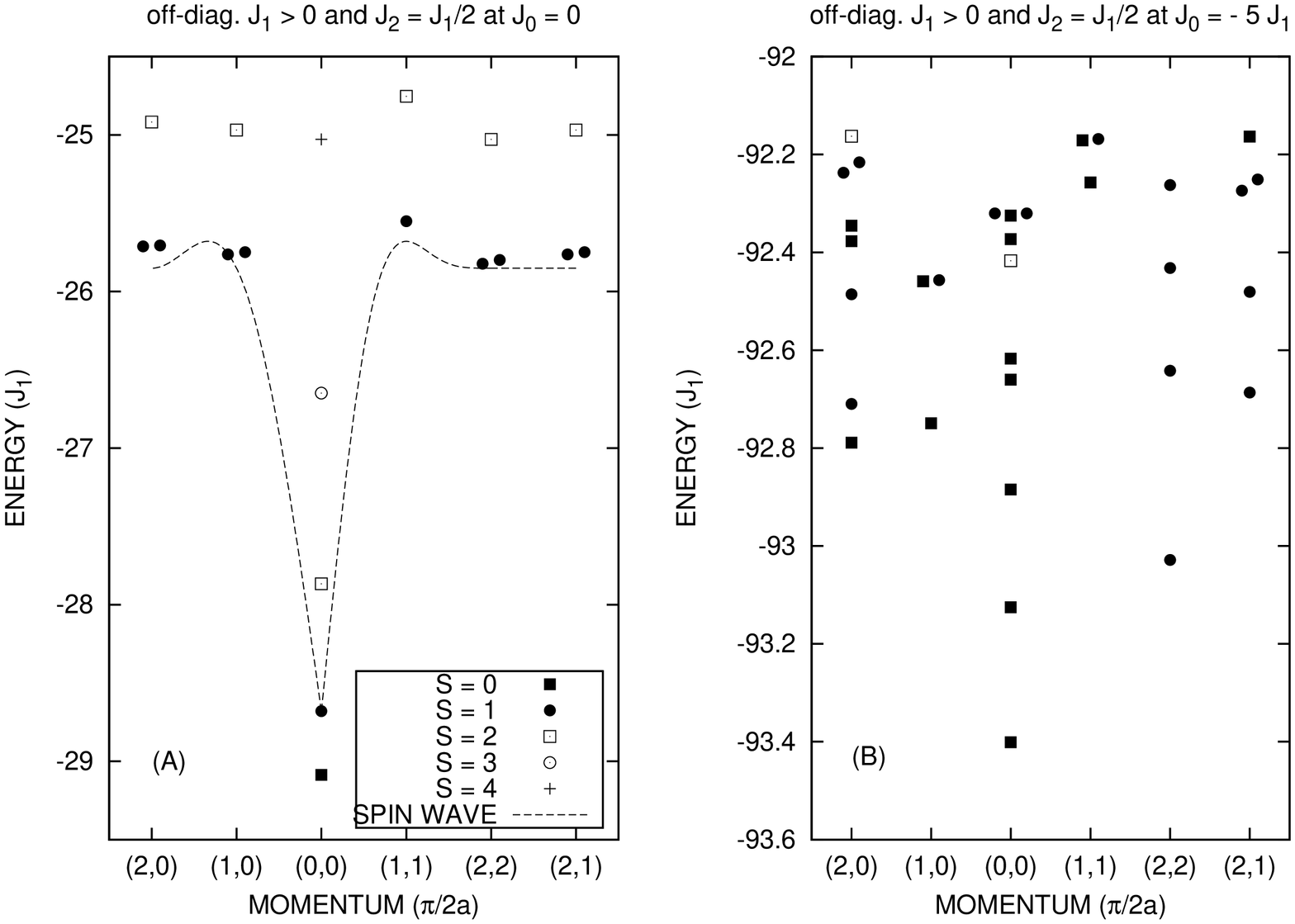}
\caption{Shown is the low-energy spectrum for 
$4\times 4\times 2$ spin-1/2 moments
that experience off-diagonal 
anti-ferromagnetic and frustrating 
Heisenberg exchange at weak and at moderately strong Hund's rule coupling.
The lowest-energy spin-1 state at momentum $(0,0)$ is used as the reference for the
linear spin-wave approximation.
}
\label{af+_spctra}
\end{figure}

\begin{figure}
\includegraphics[scale=0.70, angle=-90]{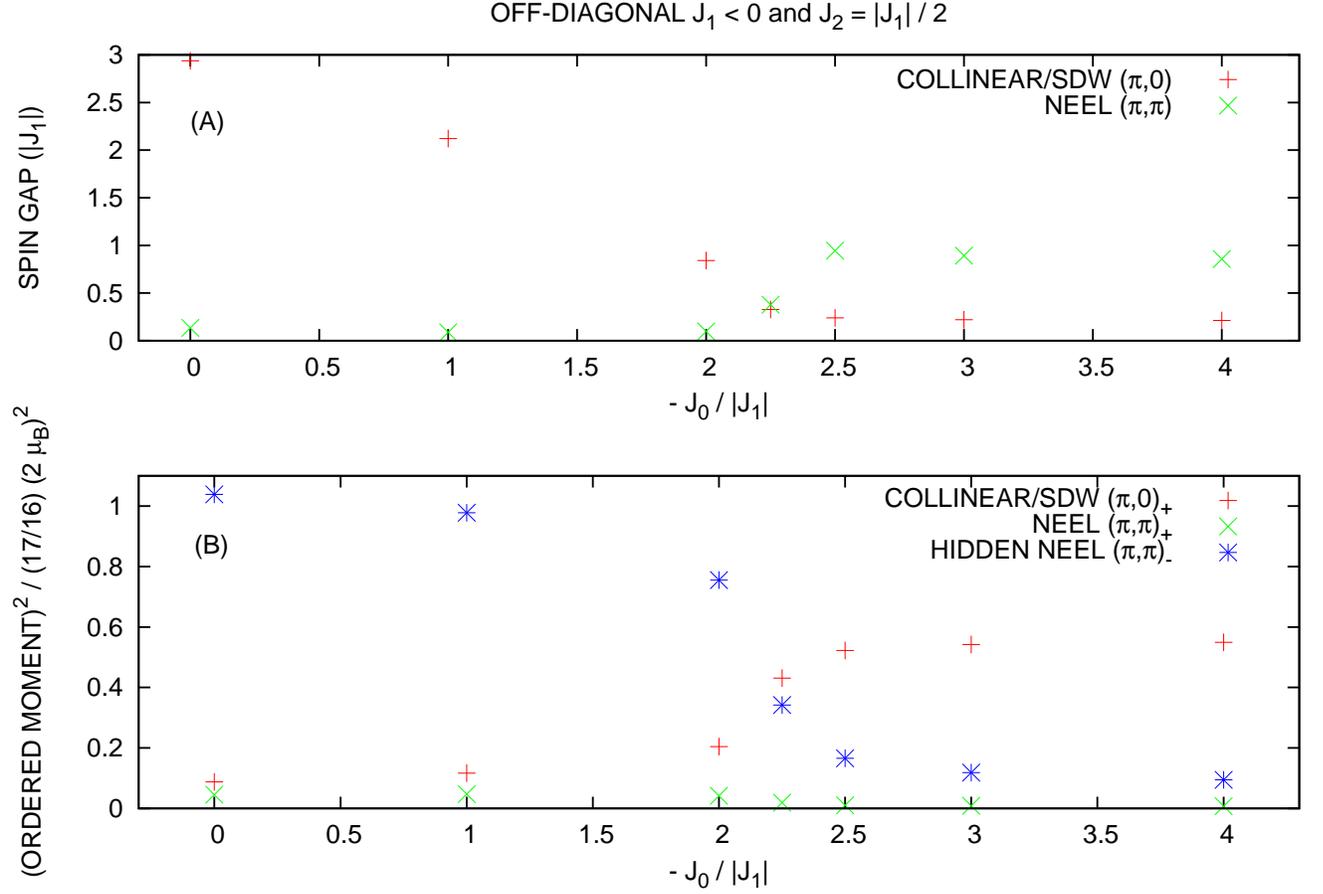}
\caption{The evolution of the spin gap with Hund's rule coupling at wave numbers that
correspond to collinear and  to N\' eel order is shown
for off-diagonal $J_1 < 0$ and off-diagonal $J_2 = |J_1| / 2$.
Also shown is the autocorrelation
$\langle {\bf O} ({\bf k})_\pm \cdot {\bf O} (-{\bf k})_\pm \rangle_0$
 of the order parameter (\ref{OP}) for true (+) and for hidden (-)
magnetic order as a function of Hund's rule coupling.
It is normalized to its value in the true ferromagnetic state.}
\label{fe+_gaps_order}
\end{figure}

\begin{figure}
\includegraphics[scale=0.70, angle=-90]{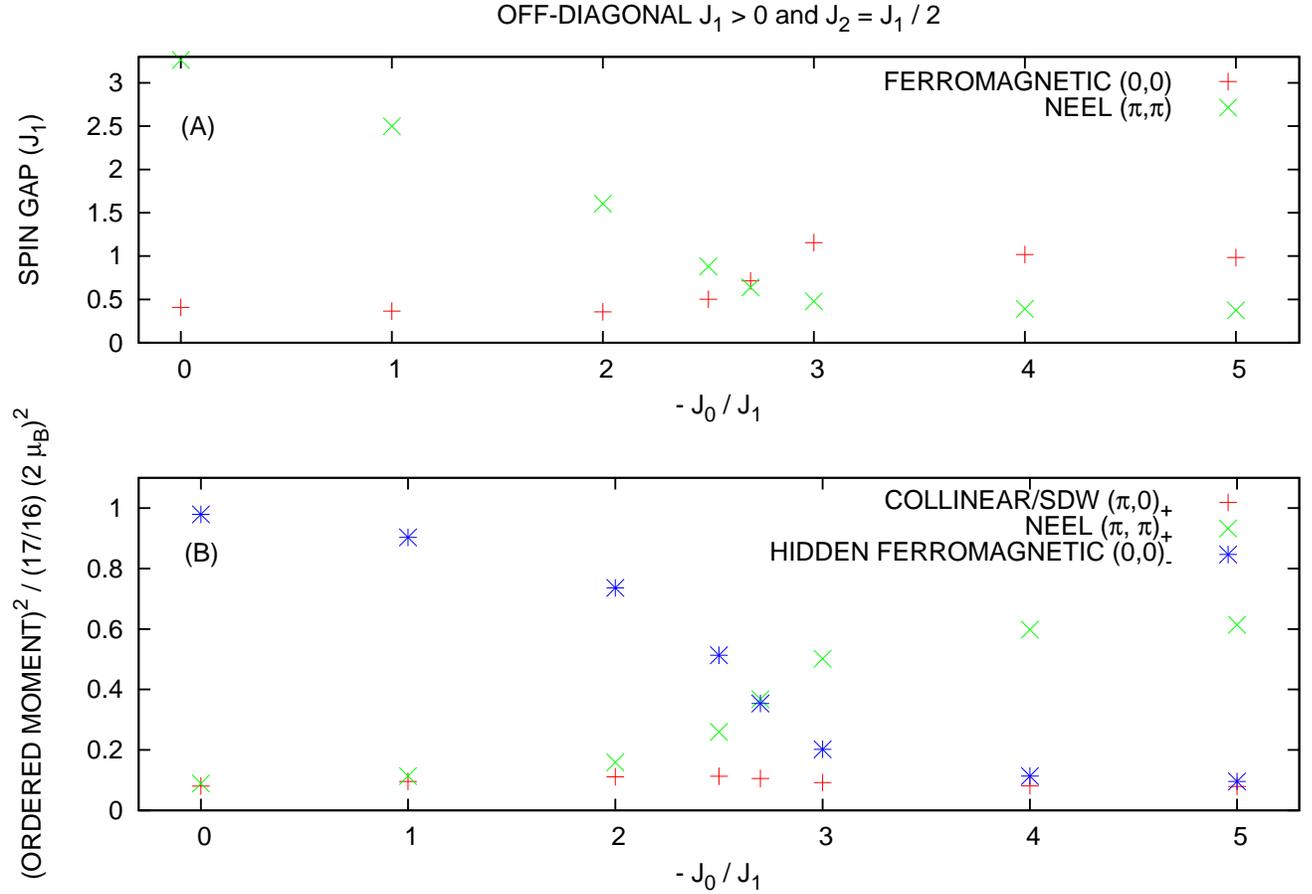}
\caption{The spin gap at wavenumbers corresponding to ferromagnetic and to N\' eel order
are displayed as a function of Hund's rule coupling 
for off-diagonal $J_1 > 0$ and off-diagonal $J_2 = J_1 / 2$.  
Also displayed is the autocorrelation of the order parameter (\ref{OP}) 
versus Hund's rule coupling.
It is normalized to its value in the true ferromagnetic state.}
\label{af+_gaps_order}
\end{figure}

Figures \ref{fe+_spctra} and \ref{af+_spctra}
show how the low-energy spectrum of the $J_0$-$J_1$-$J_2$ model  (\ref{j0j1j2})
evolves with the strength of the Hund's rule coupling 
in the case of maximum off-diagonal frustration:
$J_1^{\alpha,\alpha} = 0 = J_2^{\alpha,\alpha}$, 
while $J_1^{\alpha,\beta} = J_1$ and $J_2^{\alpha,\beta} = |J_1|/2$ for $\alpha \neq \beta$.
Respectively, they correspond to 
ferromagnetic and to antiferromagnetic  nearest-neighbor Heisenberg exchange,
$J_1 < 0$ and 
$J_1 > 0$.
Notice first the coincidence at weak Hund's rule coupling, $J_0 = 0$,
between the previous linear spin-wave  approximation
about hidden-order 
shown in  fig. \ref{sw}
with the present exact-diagonalization results.
It suggests
that long-range hidden magnetic order indeed exists.
Second, notice that the lowest energy spin-1 excitation is not the first
but the second excited
state at strong Hund's rule coupling.  
This suggests
that a nonzero spin gap exists at maximum magnetic frustration.
We have checked that the low-energy spectrum of the 
corresponding  $J_1$-$J_2$ model at spin $s = 1$ is very similar
by setting $J_{1 (2)}^{\alpha,\beta} = J_{1 (2)}$.
Both a spin-wave analysis  at large spin $s$ \cite{j1j2_sw}
and series-expansion studies\cite{j1j2_se} at $s = 1/2$
find a spin gap at maximum frustration
for the $J_1$-$J_2$ model.
The spin gap then likely persists at $s = 1$,
which argues in favor of a spin gap in the present off-diagonal case.
Both sets of spectra are then
consistent with a transition from a magnetically frustrated state that shows a spin gap,
but that obeys Hund's rule,
to an unfrustrated hidden-order state that violates Hund's rule.
Figures \ref{fe+_gaps_order} (A) and \ref{af+_gaps_order} (A)
display level crossings of the lowest-energy spin excitations,
which are consistent with 
such a quantum phase transition.
It can be shown that a transition 
into  hidden magnetic order
from true magnetic order
of collinear or of N\' eel type
is expected at $J_0 = - 2 |J_1|$
for maximum frustration, $J_2 = |J_1|/2$,
in the classical limit at large spin $s$.

Figures  \ref{fe+_gaps_order} (B)  and \ref{af+_gaps_order} (B)
also show the evolution of relevant magnetic order parameters with Hund's rule coupling.
They further confirm the interpretation that a quantum phase transition into hidden order
takes place near $J_0 = - 2 |J_1|$.
The ordered moment is obtained here by computing the autocorrelation
$\langle {\bf O} ({\bf k})_\pm \cdot {\bf O} (-{\bf k})_\pm \rangle_0$
of the order parameter
\begin{equation}
{\bf O} ({\bf k})_\pm = \sum_i e^{i{\bf k}\cdot{\bf r}_i} [{\bf S}_i(1) \pm {\bf S}_i(2)]
\label{OP}
\end{equation}
over the groundstate.
Figure \ref{fe+_gaps_order} (B), in particular,
displays how the square of the ordered moment for true collinear/SDW order (+)
decays once the system transits into hidden order at off-diagonal $J_1 < 0$. 
Figure \ref{af+_gaps_order} (B) displays
how the same occurs for true N\' eel order ($\times$) at off-diagonal $J_1 > 0$.
The former is notably consistent with the low ordered moment that is observed
by elastic neutron diffraction in an undoped parent compound to the 
recently discovered Fe-As high-$T_c$ superconductors\cite{delacruz}.
It must be emphasized, however, that the low-energy spectrum of hidden order 
contains observable spin-wave excitations
with energies that collapse to the groundstate energy
either at the N\' eel wave number $(\pi,\pi)$,
or at the ferromagnetic wave number $(0,0)$.
(See figs. \ref{sw} - \ref{af+_spctra}.)

\begin{figure}
\includegraphics[scale=0.70, angle=-90]{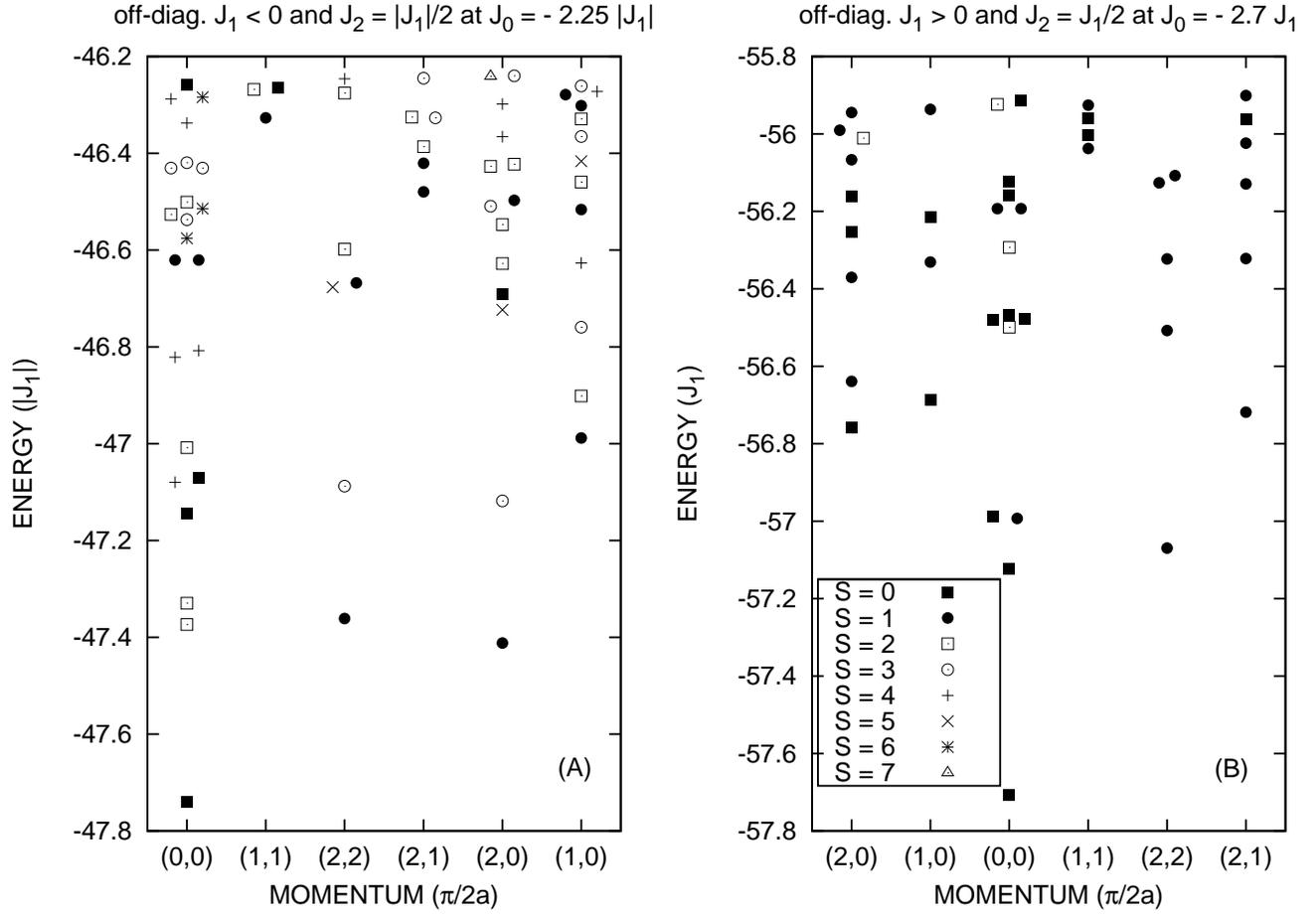}
\caption{Shown is the low energy spectrum
 for $4\times 4\times 2$ spin-1/2 moments that experience maximum
off-diagonal frustration near the transition point into hidden order.}
\label{cp_spctra}
\end{figure}

Recent inelastic neutron scattering measurements on 
an undoped parent compound of Fe-As superconductors
find a small spin gap at the collinear/SDW wave number $(\pi,0)$
on the other hand\cite{sw_exp}.
Figures \ref{fe+_gaps_order} and \ref{cp_spctra} (A) are consistent with 
both a reduced moment for collinear/SDW order
and a small spin gap at $(\pi,0)$ at the transition into hidden N\' eel  order
for off-diagonal $J_1 < 0$.
The undoped parent compounds of Fe-As superconductors could then lie {\it at} the
transition point into hidden magnetic order.

In conclusion, we have identified a route to low ordered magnetic moments in the undoped parent
compounds of the recently discovered Fe-As superconductors
that is based on the weakening of Hund's
rule by frustrating Heisenberg exchange interactions between different $3d$ orbitals 
in neighboring iron atoms.  We predict, however, that the spin-wave  excitation energy 
vanishes either at the wave number $(\pi,\pi)$ or at the wave number $(0,0)$
deep inside the respective
hidden-order phases where Hund's rule is violated.  (Cf. ref. \cite{sw_exp}.)

The authors thank Radi Al Jishi and Zlatko Tesanovic for discussions. 
Exact diagonalization of the $J_0$-$J_1$-$J_2$
model (\ref{j0j1j2}) was carried out on the SGI Altix 4700
at the 
AFRL DoD Supercomputer Resource Center.
This work was supported in part by the US Air Force
Office of Scientific Research under grant no. FA9550-06-1-0479 (JPR)
and by the National Science Foundation under grant no. DMR-0606566 (EHR).

\end{document}